\documentclass[11pt]{article}
\usepackage{amssymb,amsmath,latexsym,amscd,amsfonts}
\usepackage{graphics}
\usepackage{epsfig}

\newtheorem{thm}{Theorem}[section]

\newtheorem{definition}{Definition}[section]

\newtheorem{lem}{Lemma}[section]

\newtheorem{cor}{Corollary}[section]

\newenvironment{proof}{{\noindent{\bf Proof: }}}{$\hfill\Box$}

\newcommand{\ignore}[1]{}

\newcommand{\onote}[1]{}
\newcommand{\dfinalnote}[1]{}
\newcommand{\znote}[1]{}

\def\hpic #1 #2 {\mbox{$\begin{array}[c]{l}
\epsfig{file=#1,height=#2} \end{array}$}}
\def\vpic #1 #2 {\mbox{$\begin{array}[c]{l}
\epsfig{file=#1,width=#2}\end{array}$}}

\begin{document}
\title{ An Efficient Algorithm to Recognize Locally Equivalent Graphs in Non-Binary Case}


\author{{Mohsen Bahramgiri} \thanks{Massachusetts Institute of Technology,
Mathematics department \& Computer Science and Artificial
Intelligence Laboratory,\ m\underline{ }bahram@mit.edu.}
 \and Salman Beigi\thanks{Massachusetts Institute of Technology,
Mathematics department,\ salman@math.mit.edu.} }

\date{February 09, 2007}

\maketitle{}

\begin{abstract} Let $v$ be a vertex of a graph $G$.
By the local complementation of $G$ at $v$ we mean to complement
the subgraph induced by the neighbors of $v$. This operator can be
generalized as follows. Assume that, each edge of $G$ has a label
in the finite field $\mathbf{F}_q$. Let $(g_{ij})$ be set of
labels ($g_{ij}$ is the label of edge $ij$). We define two types
of operators. For the first one, let $v$ be a vertex of $G$ and
$a\in \mathbf{F}_q$, and obtain the graph with labels
$g'_{ij}=g_{ij}+ag_{vi}g_{vj}$. For the second, if $0\neq b\in
\mathbf{F}_q$ the resulted graph is a graph with labels
$g''_{vi}=bg_{vi}$ and $g''_{ij}=g_{ij}$, for $i,j$ unequal to
$v$. It is clear that if the field is binary, the operators are
just local complementations that we described.

The problem of whether two graphs are equivalent under local
complementations has been studied, \cite{bouchalg}. Here we
consider the general case and assuming that $q$ is odd, present
the first known efficient algorithm to verify whether two graphs
are locally equivalent or not.

\end{abstract}

\section{Introduction}

A {\it labeled graph} is a graph all of whose edges have a label
chosen from a (finite) field. This definition covers the usual
graphs when one restricts the filed to the binary field,
$\mathbf{F}_2$. We want to define the notion of {\it local
equivalency} over (labeled) graphs but, for simplicity, let us
first consider the binary case. In the binary case, i.e., when
the field is $\mathbf{F}_2$, consider the following operation,
called {\it local complementation}. Choose a vertex, and replace
the subgraph induced on the neighbors of this vertex by its
complement. Two graphs are called {\it locally equivalent} if one
can obtain one of them from the other by applying some local
operations as above.

In general, when the field is not binary, two types of operators
are involved. The first one is just the generalized version of the
operator in the binary case. Let the graph $G$ be labeled with
labels forming a symmetric matrix $G=(g_{ij})$ with zero diagonal
over $\mathbf{F}_q$, where $q$ is a power of an odd prime number,
and $\mathbf{F}_q$ is the field with $q$ elements. Let $v$ be a
vertex of this graph, and $a\in \mathbf{F}_q$. We define
$G*_{a}v$ to be the graph with labels $G'=(g'_{ij})$, where
$g'_{ij}=g_{ij}+a\,g_{vi}g_{vj}.$ In the second type of operators
we multiply the edges incident to a vertex $v$ by a non-zero $b
\in \mathbf{F}_q$, and denote this graph by $G\circ_{b}v$. In
other words, $G\circ_{b}v$ is the graph with labels
$G''=(g''_{ij})$, where $g''_{vi}=bg_{vi}$ and $g''_{ij}=g_{ij}$
for $i, j$ unequal to $v$. Similar to the previous situation, two
graphs are called locally equivalent if one of them can be
obtained from the other by applying a series of operators $*$ and
$\circ$.

Studying and investigating the local equivalency of graphs has
become a natural problem in quantum computing, and playing a
significant role especially in {\it quantum error correcting
codes}, due to the recent work of \cite{knill}, \cite{our},
\cite{moor1} and \cite{marc} . Namely, in the quantum computing
setting, some states, called {\it graph states}, have a
description as the common eigenvectors of a subgroup of the {\it
Pauli group}. These states are called graphs states because their
associate subgroup is defined bases on a labeled graphs. Using
graph states, we may be able to create more preferable {\it
quantum codes}, due to the property that the obtained codes have
relatively shorter descriptions, and are more algebraically
structured. Hence, combining the theory of quantum error
correcting codes and the tools in graph theory, leads us to
describe and investigate the properties of graph states more and
more deeply.

The key point is that, we can obtain one graph state from another
by applying elements of, what is called {\it local Clifford
group}. If two states are equivalent under local Clifford group,
they present similar properties in quantum computing. In fact, as
shown in \cite{our} and \cite{moor1}, two graph states are
equivalent under the local Clifford group if their associated
graphs are locally equivalent by the local operators described
earlier. So, this question is coming up naturally that, when two
graphs are equivalent up to these operators, and how we can
recognize them.

The special case of $q=2$, has been studied in the work of
$Andr\acute{e}\ Bouchet$, \cite{bouchtree}, \cite{bouchet}, and a
polynomial time algorithm for recognizing the local equivalency of
two (simple) graphs is described in \cite{bouchalg}. In fact, he
showed that, for any two graphs there is a system of equations
such that, the two graphs are locally equivalent iff those
equations have a solution.

Same as binary case, when $q$ is odd, recognizing the locally
equivalent graphs is equivalent to solving a system of equations,
some of which are linear and the rest are quadratic. But, their
algebraic structure are different, and is more non-linear compared
to the binary case. Indeed, in the binary field every element
satisfies $a^2=a$, and hence quadratic equations on binary field
exhibit linear properties. The algorithm described in
\cite{bouchalg} takes the advantage of this property of
$\mathbf{F}_2$. The situation in non-binary case is completely
different, and the quadratic equations do not exhibit linear
properties in general. In the present paper, we study in details
the structure of the solutions of these equations, and present an
efficient algorithm to solve the problem of recognizing locally
equivalent graphs.

\subsection{ Main ideas}

The main ideas in this paper are as follows. First of all, we
introduce {\it isotropic systems} that are geometrically known
objects, and define an equivalency relation on them. In this
definition two isotropic systems are equivalent if a system of
equations has a solution. Then, we define the isotropic system
associated to a (labeled) graph, and show that two graphs are
locally equivalent if their associated isotropic systems are
locally equivalent. Using this idea, we convert the problem of
local equivalency of graphs to the existence of a solution for a
system of algebraic equations.

Unfortunately, these equations are not all linear. So in general,
it is hard to decide whether there is a solution or not. But, in
our case, their solutions have some nice properties. In fact, we
prove that, if there is one solution then there are many. In
other words, if two graphs are locally equivalent then, in some
sense, there are many solutions for their associated system of
equations.

The idea to prove this property, is to correspond the solutions of
the system of equations for two graphs, to the solutions to some
equations associated to just one of them. What we call them {\it
internal solutions}. In fact, we show that, instead of studying
local operators that convert one graph to the other, it is
sufficient to know set of local operators that send a graph back
to itself. This correspondence allows us to somehow consider a
linear structure for the set of solutions, and to show that if
there is a solution then the solutions contain an affine subspace
of constant co-dimension. In other words, if there is a solution
then there are many, and so it is not hard to find one of them.

\section{Isotropic systems and locally equivalent graphs} \label{section:iso}

Assume that $p$ is an odd prime number, and $\mathbf{F}_q$ is the
field of $q$ elements with characteristic $p$. Suppose that $G$ is
a graph on $n$ vertices. We call $G$ a labeled graph on
$\mathbf{F}_q$, if labels of its edges form an $n\times n$
symmetric matrix $G=(g_{ij})$ with zero diagonal over
$\mathbf{F}_q$, where $g_{ij}$ is the label of edge $ij$.

\subsection{Local operators over graphs}

\begin{definition} Let $G$ be a labeled graph with labels
forming a symmetric matrix $G=(g_{ij})$ on $\mathbf{F}_q$. For
vertex $v$ of $G$ and $a\in \mathbf{F}_q$, define $G*_{a} v$ to be
a graph with the label matrix $G'=(g'_{ij})$ such that for all
$i$, $g'_{vi}=g_{vi}$, and for $i,j$ unequal to $v$,
$$g'_{ij}=g_{ij}+ag_{vi}g_{vj},$$ and moreover, $G'$ is symmetric with zero diagonal.

Also, for a non-zero number $b\in \mathbf{F}_q$ define $G\circ_b
v$ to be a graph with the label matrix $G'=(g'_{ij})$ such that
for all $i$, $g'_{iv}=bg_{iv}$, and $g'_{ij}=g_{ij}$ if $i,j$ are
unequal to $v$, and again, $G'$ is symmetric with zero diagonal.

\end{definition}

Two graphs $G$ and $G'$ are called {\it locally equivalent} if
there exists a sequence of above operations that acting on $G$
gives $G'$. Notice that, these operations are invertible, so that
this is an equivalency relation.

\subsection{Isotropic systems}

Let $\mathbf{F}^n_q$ be the $n$-dimensional vector space over
$\mathbf{F}_q$, and consider the standard bilinear form on it.
That is, for vectors $$X=(x_1,\cdots ,x_n),\ Y=(y_1, \cdots, y_n
),$$ in $\mathbf{F}^n_q$, define
$$\langle X,Y \rangle=\sum ^{n}_{i=1} x_iy_i.$$
$\langle \, .,\, . \rangle$ is a non-degenerate, symmetric
bilinear form. Using this form, we define a non-degenerate
anti-symmetric bilinear form on $\mathcal{V}=\mathbf{F}^{2n}_q$,
the $2n$-dimensional vector space over $\mathbf{F}_q$. For vectors
$(X, X')$ and $(Y, Y')$ in $\mathcal{V}$, set
$$\langle (X,X'),(Y,Y')\rangle = (X,X') \cdot \Lambda \cdot (Y,Y')^T,$$
where $$\Lambda=\begin{pmatrix}
  0 & I_n \\
  -I_n & 0
\end{pmatrix}.$$
In other words,
$$\langle (X, X' ), (Y, Y')\rangle= \langle X,Y' \rangle
-\langle X',Y \rangle.$$ Due to the nature of anti-symmetric
forms, we are now in the situation of introducing a geometrically
known concept, {\it isotropic systems}.

\begin{definition} A subspace $\mathcal {W}$ of $\mathcal{V}$ is called an {\it
isotropic system }, if it is an $n$-dimensional subspace and
$\langle (X, X' ), (Y, Y')\rangle=0$, for any $(X, X' ), (Y,
Y')\in \mathcal{W}$. In fact, since $dim \mathcal{W}=n$ and
$\langle .,. \rangle$ is a non-degenerate bilinear form, we have
\begin{equation}
\mathcal{W}=\{V\in \mathcal{V}: \langle V, W\rangle=0,\ \forall\,
W\in \mathcal{W}\}. \label{equ1}
\end{equation}

\end{definition}

In this paper the basic examples of isotropic systems are {\it
isotropic systems associated to graphs}. For every graph $G$, let
$\mathcal{W}_G$ be the vector subspace generated by the rows of
matrix $\begin{pmatrix}
  I & \mid & G
\end{pmatrix}$. That is
$$ \mathcal {W}_G= \{ X \cdot \begin{pmatrix}
  I & \mid & G
\end{pmatrix}:\ X\in \mathbf{F}^{n}_q\},$$
where $\begin{pmatrix}
  I & \mid & G
\end{pmatrix}$ is a matrix with two $n\times n$ blocks which the first
one is identity. Since, $G$ is a symmetric matrix
$$\begin{pmatrix} I & \mid & G \end{pmatrix} \cdot \Lambda \cdot \begin{pmatrix} G & \mid & -I
\end{pmatrix} ^T =G^T -G=0,$$
and we conclude that $\mathcal {W}_G$ forms an isotropic system
which is called the {\it isotropic system associated to $G$}.

\subsection{Locally equivalent isotropic systems}

\begin{definition} Suppose that $A$ is a $2n\times 2n$ matrix
$$A=\begin{pmatrix}
  Z & T \\
  X & Y
\end{pmatrix},$$ consisting of four diagonal matrices $X=diag(x_1, \cdots,  x_n),
Y=diag(y_1, \cdots, y_n)$, $Z=diag(z_1, \cdots, z_n)$ and
$T=diag(t_1, \cdots, t_n)$. $A$ is called {\it normal} if
$$Y\cdot Z-X\cdot T=I.$$
Notice that, every normal matrix is invertible, its inverse is
normal as well, and in addition the multiplication of normal
matrices is again normal. In particular, when $Z=I$ and $X=0$, we
call $A$ trivial.
\end{definition}

For an isotropic system $\mathcal{W}$ and a normal matrix $A$,
define $$\mathcal{W}\cdot A=\{W \cdot A: W\in \mathcal{W}\}.$$
Since $A$ is normal, one can verify that $\mathcal{W}\cdot A$ is
also an isotropic system.

Two isotropic systems $\mathcal{W}$ and $\mathcal{W'}$ are called
{\it locally equivalent} if there exists a normal matrix $A$ such
that $\mathcal{W'}=\mathcal{W}\cdot A$. By the properties of
normal matrices, it is clear that this is an equivalency relation.
The importance of this relation would be clear once we state the
following theorem.

\begin{thm}\label{leq}{ Two graphs $G$ and $H$ on the same vertex
sets are locally equivalent if and only if their associated
isotropic systems are locally equivalent.

}\end{thm}

\begin{proof}{ For the {\it only if }  part, it is sufficient to show
that the systems $\mathcal {W}_G, \mathcal {W}_{G*_a i}$, and
also the systems $\mathcal {W}_G, \mathcal {W}_{G\circ_b i}$ are
locally equivalent. First, note that the rows of $\begin{pmatrix}
  I & \mid & G*_ai
\end{pmatrix}$ form a basis for $\mathcal {W}_{G*_ai}.$ Let $$A=\begin{pmatrix}
  Z & T \\
  X & Y
\end{pmatrix},$$  where $X=diag(0, \cdots ,0,-a ,0,\cdots,0), Y=I, Z=I$ and $T=0$. We just need to check that the
rows of $\begin{pmatrix}
  I & \mid & G
\end{pmatrix} \cdot A $ are orthogonal to the rows of $\begin{pmatrix}
  I & \mid & G*_ai
\end{pmatrix}$.

Consider the $j$-th row of $\begin{pmatrix}
  I & \mid & G
\end{pmatrix} \cdot A $, and the $k$-th row of $\begin{pmatrix}
  I & \mid & G*_ai
\end{pmatrix}$. If $j\neq i,$ and $k\neq i$, the product of these two rows is
$(g_{jk}+ag_{ik}g_{ij}-ag_{ij}g_{ik})-(g_{jk})=0$. If $j=i$ and
$k\neq i$, it is $(g_{ik})-(g_{ik})=0$, and finally, if $j\neq
i$, and $k=i$ then this product is again $(g_{ij})-(g_{ij})=0$.
Also, the $i$-th row of these matrices are equal, and therefore,
according to the definition of the inner product, these rows are
orthogonal. Thus, $\mathcal {W}_G$ and $\mathcal {W}_{G*_ai}$ are
locally equivalent.

Similarly, for $\mathcal {W}_G, \mathcal {W}_{G\circ_b i}$, let
$$B=\begin{pmatrix}
  Z' & T' \\
  X' & Y'
\end{pmatrix},$$
where $X'=0$, $T'=0$ and
$$Y'=diag(1,\cdots,1, b, 1,\cdots, 1), Z'=diag(1,\cdots,1 , b^{-1}, 1,\cdots, 1).$$ Similar to the
previous case, one can easily check that the rows of
$\begin{pmatrix}
  I & \mid & G
\end{pmatrix} \cdot B $ and $\begin{pmatrix}
  I & \mid & G\circ_bi
\end{pmatrix}$ are orthogonal, and therefore, $\mathcal {W}_G, \mathcal{W}_{G\circ_b
i}$ are locally equivalent.

To prove the {\it if} part, suppose that $\mathcal{W}_G$ and
$\mathcal{W}_{H}$ are locally equivalent. Then there exists a
normal matrix $$A=\begin{pmatrix}
  Z & T \\
  X & Y
\end{pmatrix},$$ where, $$X=diag(x_1, \cdots, x_n),
Y=diag(y_1, \cdots, y_n),$$
$$Z=diag(z_1, \cdots, z_n), T=diag(t_1, \cdots, t_n),$$ and rows of $\begin{pmatrix}
  I & \mid & G
\end{pmatrix} \cdot A $ form a basis for the isotropic system
$\mathcal{W}_{H}$. Therefore, there exists an invertible matrix
$U$ such that $U\cdot \begin{pmatrix}
  I & \mid & G
\end{pmatrix} \cdot A=\begin{pmatrix}
  I & \mid & H
\end{pmatrix} $.

For every $i$, let $$A_i=\begin{pmatrix}
  Z_{i} & T_{i} \\
  X_{i} & Y_{i}
\end{pmatrix},$$ where $$Z_i=diag(1,\cdots,1, z_i,1,\cdots ,1),
T_i=diag(0,\cdots,0, t_i,0,\cdots, 0),$$ $$X_i=diag(0,\cdots,
0,x_i,0,\cdots, 0),  Y_i=diag(1, \cdots,1, y_i,1,\cdots, 1).$$ The
matrices $A_i$'s all commute and $A=A_1A_2\cdots A_n$.

We prove the theorem by induction on the number of non-trivial
matrices $A_i$. If all $A_i$'s are trivial then $A$ is also
trivial, and we have $$\begin{pmatrix}
  I & \mid & G
\end{pmatrix}\cdot A=\begin{pmatrix}
  I & \mid & G
\end{pmatrix}\cdot \begin{pmatrix}
  I & T \\
  0 & Y
\end{pmatrix}=\begin{pmatrix}
  I & \mid & T+GY
\end{pmatrix}.$$  Therefore, we have $U\cdot \begin{pmatrix}
  I & \mid & T+GY
\end{pmatrix}=\begin{pmatrix}
  I & \mid & H
\end{pmatrix}$. Looking at the first blocks in this equation, we get that $U=I$, and $T+GY=H$.
The diagonal of $H$ is zero, and hence $T=0$. Also, $A$ is a
normal matrix, so that, $A=I$ and $G=H$.

Hence, suppose that at least one of $A_i$'s is non-trivial. We
consider two cases:

\vspace{4mm}

\noindent {\bf Case(i).} $z_{i_0}\neq 0$ for some $i_0$, where
$A_{i_0}$ is non-trivial. Let $\begin{pmatrix}
  I & \mid & G
\end{pmatrix}A_{i_0}=\begin{pmatrix}
  V & \mid & D
\end{pmatrix}$. Then,

$$V=\begin{pmatrix}
  1      & 0  & \cdots & x_{i_0}g_{1i_0}& \cdots & 0      &   \\
  0      & 1  &        &  \vdots         &        & 0      &   \\
  \vdots &    & \ddots &  \vdots         &        & \vdots &   \\
  \vdots &    &        & z_{i_0}         &        & \vdots &   \\
  \vdots &    &        &  \vdots         & \ddots & 0      &   \\
  0      & 0  & \cdots & x_{i_0}g_{ni_0}& \cdots & 1      &
\end{pmatrix}.$$
In order to get to the inverse of $V$, we should multiply the
$i_0$-th row by $z^{-1}_{i_0}$, and then by $-x_{i_0}g_{ji_0}$ and
add it to the $j$-th row, for any $j\neq i_0$. Thus,
$V^{-1}\begin{pmatrix}
  I & \mid & G
\end{pmatrix}A_{i_0}=\begin{pmatrix}
  I & \mid & V^{-1}D
\end{pmatrix}$, and the $jk$-th entry of $V^{-1}D$, for $j$ unequal to $k,$ and $i_0$, is
$$(V^{-1}D)_{jk}=g_{jk}-z^{-1}_{i_0}x_{i_0}g_{i_0j}g_{i_0k}.$$ Also, for $j\neq i_0$,
$$(V^{-1}D)_{ji_0}=(V^{-1}D)_{i_0j}=z^{-1}_{i_0}g_{i_0j}.$$
The matrix $V^{-1}D$ may have non-zero entries on its diagonal.
But, by elementary linear algebra, there exists a trivial matrix
$A'$, such that all the entries of $V^{-1}\begin{pmatrix}
  I & \mid & G
\end{pmatrix}A_{i_0}A'$ are equal to $V^{-1}\begin{pmatrix}
  I & \mid & G
\end{pmatrix}A_{i_0}$
except those, on the diagonal of the second block, which are zero.
Therefore, the rows of $V^{-1}\begin{pmatrix}
  I & \mid & G
\end{pmatrix}A_{i_0}A'$ span an isotropic system associated to some graph,
and by the above equalities, this graph is nothing but
$G*_{(-z^{-1}_{i_0}x_{i_0})}i_0 \circ_{(z^{-1}_{i_0})}i_0.$

On the other hand, we have $$\begin{pmatrix}
  I & \mid & H
\end{pmatrix}=U\begin{pmatrix}
  I & \mid & G
\end{pmatrix}A=UV(V^{-1}\begin{pmatrix}
  I & \mid & G
\end{pmatrix}A_{i_0}A')(A'^{-1}A''),$$ where, $A''$
is equal to the multiplication of all $A_j$'s, except $A_{i_0}$.

Now, $V^{-1}\begin{pmatrix}
  I & \mid & G
\end{pmatrix}A_{i_0}A'$ is an isotropic system, associated to the graph $$G*_{(-z^{-1}
_{i_0}x_{i_0})}i_0 \circ_{z_{i_0}}i_0.$$ Also, the number of
non-trivial terms in $A'^{-1}A''$ is strictly less than the
number of non-trivial terms in $A$, and therefore, by induction,
we obtain the desired result.

\vspace{4mm}

\noindent{\bf Case(ii).} $z_{i}= 0$ for all $i$'s, where $A_{i}$
is non-trivial. Notice that, in this case $x_i\neq 0$ for any $i$,
where $A_i$ is non-trivial. Because, $A$ is a normal matrix and
$y_iz_i-x_it_i=1$. Suppose that, $A_{i_0}$ is non-trivial. If for
every non-trivial $A_j$, $g_{i_{0}j}= 0$, then the $i_0$-th row of
the first block of $\begin{pmatrix}
  I & \mid & G
\end{pmatrix}A$
is zero. Hence, it is not invertible and the first block of
$U\begin{pmatrix}
  I & \mid & G
\end{pmatrix}A$
can not be identity. Thus, there exists an $i_1$, such that
$A_{i_1}$ is non-trivial and $g_{i_{0}i_1}\neq 0$. Therefore, the
first block of $\begin{pmatrix}
  I & \mid & G
\end{pmatrix}A_{i_0}A_{i_1}$ is
$$V=\begin{pmatrix}
  1      &\cdots & x_{i_0}g_{1i_0}   & x_{i_1}g_{1i_1}  & \cdots & 0      &   \\
  0      &\ddots & \vdots             &  \vdots           &        & 0      &   \\
  \vdots &       & 0                  & x_{i_1}g_{i_0i_1}&        & \vdots &   \\
  \vdots &       & x_{i_0}g_{i_1i_0} &    0              &        & \vdots &   \\
  0      &       & \vdots             &  \vdots           & \ddots & 0      &   \\
  0      &\cdots & x_{i_0}g_{ni_0}   & x_{i_1}g_{ni_1}  & \cdots & 1      &
\end{pmatrix}.$$

In order to invert $V$, one has to multiply the $i_0$-th and the
$i_1$-st rows by $(x_{i_1}g_{i_0i_1})^{-1}$ and
$(x_{i_0}g_{i_1i_0})^{-1}$ respectively, and then multiply the
row $i_1$ by $-x_{i_0}g_{ji_0}$ and add it to the $j$-th row, for
any $j$. Also, exactly the same process for row $i_0$ must be
done. After this, we should change the rows $i_0$ and $i_1$. By
this process, we get to a matrix with identity in the first block
and a symmetric matrix on the second block. But, the diagonal
entries of this block may be non-zero. In order to handle this
issue, by multiplying by an appropriate trivial matrix $A'$, we
get
$$V^{-1}\begin{pmatrix}
  I & \mid & G
\end{pmatrix}A_{i_0}A_{i_1}A'=\begin{pmatrix}
  I & \mid & G'
\end{pmatrix},$$ where $G'=(g'_{jk})$ is the graph with entries
$$g'_{jk}=g_{jk}-g_{i_0i_1}^{-1}g_{i_0j}g_{i_1k}-g_{i_0i_1}^{-1}g_{i_1j}g_{i_0k},$$
for all $j,k\neq i_0,i_1$, and
$$g'_{i_0j}=x^{-1}_{i_0}g^{-1}_{i_0i_1}g_{i_1j},$$
$$g'_{i_1j}=x^{-1}_{i_1}g^{-1}_{i_0i_1}g_{i_0j},$$ for all $j\neq
i_0, i_1$. Moreover
$$g_{i_0i_1}=-x^{-1}_{i_0}x^{-1}_{i_1}g^{-1}_{i_0i_1}.$$
Therefore,
$$G'=G\circ_{(-g^{-1}_{i_0i_1})}i_0*_{1}i_0*_{(-1)}i_1*_{1}i_0
\circ_{(x^{-1}_{i_0}g^{-1}_{i_0i_1})}i_0\circ_{(x^{-1}_{i_1})}i_1.$$

We have $\begin{pmatrix}
  I & \mid & H
\end{pmatrix}=UV\begin{pmatrix}
  I & \mid & G'
\end{pmatrix}A'^{-1}A''$, where $A''$ is equal to the
multiplication of all $A_j$'s, except $A_{i_0}$ and $A_{i_1}$.
Also the number of non-trivial terms in $A'^{-1}A''$ is strictly
less than this number in $A$, and therefore, by induction, the
result is proved.

}\end{proof}

\section{System of equations and normal matrices}\label{section:sys}

Let $G$ and $H$ be two graphs, and $g$, $h$ be their neighborhood
functions respectively, meaning that $g(i)$ is a vector such that
its $j$-th coordinate, $g_{ij}$, is the label of the edge $ij$ in
$G$. The same holds for the graph $H$ and the function $h$. Using
theorem \ref{leq} and relation (\ref{equ1}) in the definition of
isotropic systems, one obtains that $G$ and $H$ are locally
equivalent if there exists a normal matrix
$$A=\begin{pmatrix}
  Z & T \\
  X & Y
\end{pmatrix},$$ such that rows of $\begin{pmatrix}
  I & \mid & G
\end{pmatrix} \cdot A $ are all orthogonal to the rows of  $\begin{pmatrix}
  I & \mid & H
\end{pmatrix} $. This condition is equivalent
to the following:
\begin{equation}
\langle X, g(i)\times h(j)\rangle-\langle Y, g(i)\times
e_j\rangle+\langle Z, e_i\times h(j)\rangle-\langle T, e_i\times
e_j\rangle=0, \label{equ2}
\end{equation}
for any two vertices $i, j$. Here, by $u \times v$ for two vectors
$u$ and $v$ of the same size, we mean a vector of the same size
whose $k$-th coordinate is the product of the $k$-th coordinates
of $u$ and $v$. Also, $e_i$ is a vector whose all coordinates are
zero except the $i$-th, which is one. Notice that, in this
formula, and also later on, we look at the diagonal $n \times n$
matrices as vectors of size $n$ when necessary.

Let $\mathbf{F}^{4n}_q$ be the space of the vectors of the form
$(X, Y, Z, T)$, provided with the following symmetric bilinear
form

$$\big\langle(X, Y, Z,
T),(X', Y', Z', T') \big\rangle=\langle X, X' \rangle- \langle Y,
Y' \rangle+ \langle Z, Z' \rangle -\langle T, T' \rangle.$$
Moreover, for any pair of vertices $i, j$, we define the
following function, so called {\it Lambda Function}, which plays
a key role in the whole section.
$$\lambda (i, j)=\Big(g(i)\times h(j), g(i)\times e_j, e_i \times h(j), e_i \times e_j\Big),$$
and $$\lambda (G ,H)=Span \{\lambda (i, j):\  i, j\}.$$ Notice
that, equation (\ref{equ2}) is equivalent to say that
$\lambda(i,j)$ is orthogonal to $(X, Y, Z, T)$. Therefore, the
problem of local equivalency of $G$ and $H$ reduces to the
following:

\vspace{3mm}

{\it Two graphs $G$ and $H$ are locally equivalent iff there
exists a vector $(X,Y,Z,T)$ orthogonal to $\lambda (G,H)$, and
$Y\times Z-X\times T=I.$}

\vspace{3 mm}

\subsection{Changes of $\lambda(G,H)$ under local operators}

We develop some more technical tools to attack this problem. For
$\phi=(X, Y, Z, T)$ $\in \mathbf{F}^{4n}_q$, let
$$\phi^1=(-Z, -T, -X, -Y),$$ $$\phi^2=(Y, X, T, Z),$$ and for
$\alpha \in \mathbf{F}^{4n}_q $, $a\in \mathbf{F}_q$ and $l=1,2$
define
$$\phi*_{l,a} \alpha=\phi-a \phi^l \times \alpha.$$
Also, for any subspace $N$ of $\mathbf{F}^{4n}_q$, let
$$N*_{l,a}\alpha=\{\phi*_{l,a}\alpha: \phi \in N \}.$$

\begin{lem}\label{a1}{ Let $\alpha \in \mathbf{F}^{4n}_q$ be such that $\alpha \times
\alpha^l=0$ for $l=1,2$. For $\phi , \psi \in \mathbf{F}^{4n}_q$,
and any subspace $N$ of $\mathbf{F}^{4n}_q$, the following
properties hold:
\begin{itemize}
 \item [${\rm (i)}$] $\phi \rightarrow \phi*_{l,a}\alpha $ is bijective.
\item [${\rm (ii)}$] $\langle \phi*_{l,a}\alpha, \psi*_{l,a}\alpha^l\rangle=\langle \phi, \psi \rangle.$
\item [${\rm (iii)}$] $(N*_{l,a}\alpha)^{\perp}=N^{\perp}*_{l,a}
\alpha^l$.

\end{itemize}

 }\end{lem}

\begin{proof}{ For ${\rm (i)}$, by a simple induction one can check
that after $k$ times iteration, we get
$\phi*_{l,a}\alpha\cdots*_{l,a}\alpha=\phi-ka\phi^{l}\times\alpha$.
Therefore, for $k=p$ we end up with the identity map, and hence
$\phi \rightarrow \phi*_{l,a}\alpha $ is a bijection.

For ${\rm (ii)}$, using the facts that $\langle \phi\times\psi'^l,
\psi^l \rangle=-\langle\phi^l\times\psi', \psi \rangle$ and
$\langle \phi, \psi\times\psi' \rangle$ $=$ $\langle
\phi\times\psi', \psi\rangle$, we have \begin{eqnarray*} \langle
\phi*_{l,a}\alpha,\psi*_{l,a}\alpha^l\rangle & = &\langle\phi,
\psi \rangle-a\langle\phi,
\psi^l\times\alpha^l\rangle-a\langle\phi^l\times\alpha,
\psi\rangle+a^2\langle\phi^l\times\alpha,
\psi^l\times\alpha^l\rangle \\
&=&\langle\phi, \psi\rangle-a\langle\phi\times\alpha^l,
\psi^l\rangle-a\langle\phi^l\times\alpha,
\psi\rangle+a^2\langle\phi^l\times\alpha\times\alpha^l,
\psi^l\rangle\\
&=&\langle\phi, \psi\rangle.
\end{eqnarray*}

${\rm (iii)}$ is a direct consequence of ${\rm (ii)}$.

}\end{proof}

\vspace{3mm}

In the next two theorems, we study the effect of local operations
on the set $\lambda(G,H)^\bot$, and observe that this set is
well-behaved under these types of operators.

\begin{thm}\label{ta}{ Let $G$ and $H$ be two graphs on the same
vertex set with neighborhood functions $g$ and $h$, respectively.
For every vertex $i$,
\begin{itemize}
\item [${\rm (i)}$] $\lambda(G*_{a}i, H)=\lambda(G, H)*_{1,a}(-g(i)\times g(i),-g(i)\times g(i), -e_i, -e_i)$.
\item [${\rm (ii)}$] $\lambda(G, H*_{a}i)=\lambda(G, H)*_{2,a}(h(i)\times h(i),e_i, h(i)\times h(i), e_i)$.
\end{itemize}
And more importantly,
\begin{itemize}
\item [${\rm (iii)}$] $\lambda(G*_{a}i, H)^{\perp}=\lambda(G, H)^{\perp}*_{1,a}(e_i, e_i, g(i)\times g(i), g(i)\times g(i)).$
\item [${\rm (iv)}$] $\lambda(G, H*_{a}i)^{\perp}=\lambda(G, H)^{\perp}*_{2,a}(e_i, h(i)\times h(i), e_i, h(i)\times h(i)).$

\end{itemize}

}\end{thm}

\begin{proof}{ The proof of ${\rm (ii)}$ is similar to that of ${\rm (i)}$.
Also, parts ${\rm (iii)}$ and ${\rm (iv)}$ are immediate
consequences of ${\rm (i)}$, ${\rm (ii)}$ and the third part of
lemma \ref{a1}. Hence, we just need to prove the first part.

Suppose that $g'$ is the neighborhood function of $G*_{a}i$. Then,
we have $$g'(j)=g(j)+ag_{ij}g(i)-ag^2_{ij}e_j.$$ The space
$\lambda(G*_{a}i, H)$ is generated by $\lambda'(j,k)$'s, where

\begin{eqnarray*}\lambda'(j,k)&=&\Big(g'(j)\times h(k), g'(j)\times e_k,
e_j\times h(k), e_j\times e_k\Big)\\
&=&\lambda(j,k)+ag_{ij}\Big(g(i)\times h(k), g(i)\times
e_k,0,0\Big)-ag^2_{ij} \Big(e_j\times h(k), e_j\times
e_k,0,0\Big)\\
&=&\lambda(j,k)+ag_{ij}\Big(\lambda(i,k)-(0,0,e_i\times h(k),
e_i\times e_k) \Big)\\
& &-ag^2_{ij}\Big(e_j\times h(k), e_j\times e_k,0,0\Big).
\end{eqnarray*}

Set $\lambda''(j,k)=\lambda'(j,k)-ag_{ij}\lambda'(i,k)$. It is
easy to check that $\lambda''(j,k)$'s also generate the space
$\lambda(G*_{a}i, H)$, and we have
%

\begin{eqnarray*} \lambda''(j,k)&=&\lambda(j,k)+ag_{ij}\lambda(i,k)-ag_{ij}\Big(0,0,e_i\times h(k),e_i\times
e_k\Big)\\
& & -ag^2_{ij}\Big(e_j\times h(k), e_j\times
e_k,0,0\Big)-ag_{ij}\lambda(i,k)\\
& = &\lambda(j,k)-a\Big(0,0,g(j)\times e_i\times h(k), g(j)\times
e_i \times e_k\Big)\\
& & -a\Big(g(i)\times g(i)\times e_j\times h(k),g(i)\times
g(i)\times e_j\times e_k,0,0\Big) \\
&=&\lambda(j,k)-a\Big(e_j\times h(k), e_j\times e_k, g(j)\times
h(k), g(j)\times e_k\Big)\\
& & \hspace{1.8cm} \times \Big(g(i)\times g(i), g(i)\times
g(i), e_i, e_i\Big)\\
&=&\lambda(j,k)*_{1,a}\Big(-g(i)\times g(i),-g(i)\times g(i),
-e_i, -e_i\Big).
\end{eqnarray*}
Therefore, $\lambda(G*_{a}i, H)=\lambda(G, H)*_{1,a}(-g(i)\times
g(i),-g(i)\times g(i), -e_i, -e_i)$.

}\end{proof}

To state the next theorem define
$$f_{b,i}=I+(b-1)e_i=(1,\cdots,1, b, 1,\cdots, 1),$$
for every $0\neq b\in \mathbf{F}_q$ and vertex $i$.

\begin{thm}\label{tb}{ Let $G$ and $H$ be two graphs on the same
vertex sets, and with neighborhood functions $g$ and $h$,
respectively. For every vertex $i$,
\begin{itemize}
\item [${\rm (i)}$] The map $\phi \rightarrow \phi\times (f_{b_{1},i}\ , f_{b_{2},i}\ ,f_{b_3,i}\ ,
f_{b_4,i})$ is bijective, for non-zero elements $b_1, b_2, b_3,
b_4\in \mathbf{F}_q$.
\item [${\rm (ii)}$] $\lambda(G\circ_b i, H)^{\perp}=\lambda(G,H)^{\perp}\times (f_{b^{-1},i}\ , f_{b^{-1},i}\ ,f_{b,i}\ ,
f_{b,i}).$
\item [${\rm (iii)}$] $\lambda(G, H\circ_b i)^{\perp}=\lambda(G, H)^{\perp}\times (f_{b^{-1},i}\ , f_{b,i}\ ,f_{b^{-1},i}\ ,
f_{b,i}).$

\end{itemize}

}\end{thm}

\begin{proof}{ The proof of ${\rm (i)}$ is straight forward. To prove ${\rm (ii)}$, notice that
$g'$, the neighborhood function of $G\circ_b i$, is given by
$g'(j)=g(j)\times f_{b,i}$ if $j\neq i$, and $g'(i)=bg(i)$.
Hence, if $\langle\phi, \lambda(j,k) \rangle=0$, then
$$\langle \phi\times (f_{b^{-1},i}\ , f_{b^{-1},i}\ ,f_{b,i}\ , f_{b,i}),
\lambda'(j,k)\rangle=0,$$ where $\lambda'(j,k)=\Big(g'(j)\times
h(k), g'(j)\times e_k, e_j \times h(k), e_j \times e_k\Big)$. Part
${\rm (iii)}$ is similar to ${\rm (ii)}$.

}
\end{proof}

\subsection{Changes of determinant function}

We now define the {\it determinant function} for a vector in
$\mathbf{F}^{4n}$. For a vector $\phi=(X, Y, Z, T)$, set $$det\,
\phi=Y\!\times\! Z-\!X\!\times\! T.$$ Some straight forward
computations easily lead to the proof of the following lemma.

\begin{lem}\label{tfb}{\hfill

\begin{itemize}
\item [${\rm (i)}$] $det\, \phi= det\, \phi*_{1,a}(e_i, e_i, g(i)\times g(i), g(i)\times g(i))$.
\item [${\rm (ii)}$] $det\, \phi=det\, \phi*_{2,a}(e_i, g(i)\times g(i), e_i, g(i)\times g(i)).$
\item [${\rm (iii)}$] $\det\, \phi=det\, \phi \times (f_{b^{-1},i}, f_{b^{-1},i},f_{b,i},
f_{b,i}).$
\item[${\rm (iv)}$] $\det\, \phi=det\, \phi \times (f_{b^{-1},i}, f_{b,i},f_{b^{-1},i},
f_{b,i}).$
\end{itemize}
\hfill{$\Box$}

}\end{lem}

Note that, the functions that we see on the right hand side of
parts (i) to (iv) in this lemma, are exactly the ones appear in
theorems \ref{ta} and \ref{tb}. Hence, this lemma states that the
determinant function is invariant under the action of $*$ and
$\circ$.

\subsection{What is the new picture?}

In the new setting, the problem of verifying whether or not two
graphs $G, H$ are locally equivalent, is equivalent to finding a
vector $\phi \in \mathbf{F}^{4n}_q$ such that $\phi \in \lambda
(G, H)^{\perp}$ and $det\, \phi= I$.

To get a more convenient notation, let $\Lambda (G, H)=\lambda (G,
H)^{\perp}$, and $\sigma (G, H)$ be the set of solutions, i.e.,
vectors $\phi \in \Lambda (G, H)$ satisfying $det\, \phi =I.$
Then, we get to the following picture from theorems \ref{ta},
\ref{tb} together with lemma \ref{tfb}.

If graphs $G_1, G_2$ are locally equivalent, as well as the graphs
$H_1, H_2$, then there exists a (linear) bijection $\beta$, such
that $$\Lambda(G_2, H_2)= \beta (\Lambda (G_1, H_1 )),$$
$$\sigma(G_2, H_2)= \beta (\sigma (G_1, H_1 )).$$
Even though the function $\beta$ depends on $G_1, G_2, H_1$ and
$H_2$, but it gives us useful information on the locally
equivalent graphs. Namely, if two graphs $G$ and $H$ are locally
equivalent, then roughly speaking, the relative linear position of
$\sigma(G, H)$ inside $\Lambda (G, H)$ is exactly the same as the
relative linear position of $\sigma(G, G)$ inside $\Lambda (G,
G)$. For instance, once we prove that for every graph $G$, $\sigma
(G,G)$ is a {\it large} subset of $\Lambda (G,G)$, in the sense
that it contains a linear subspace of small co-dimension, then
the same must be true for $\sigma(G,H)$ inside $\Lambda (G,H)$,
when $G$ and $H$ are locally equivalent.

\section{Internal solutions}

We have shown that, for locally equivalent graphs $G_1$ and
$G_2$, and again locally equivalent graphs $H_1$ and $H_2$, there
exists a linear bijection $\beta$, such that
$$\Lambda(G_2, H_2)= \beta (\Lambda (G_1, H_1 )),$$
$$\sigma(G_2, H_2)= \beta (\sigma (G_1, H_1 )).$$

In this section and the next one, we show that $\sigma (G,G)$ is
a {\it large} subset of $\Lambda (G,G)$, in the sense that it
contains a linear subspace of co-dimension $\leq 5$.
Consequently, by the existence of bijection $\beta$, the same
must be true for $\sigma(G,H)$ inside $\Lambda(G,H)$ when $G$ and
$H$ are locally equivalent. This observation suggests us to
consider just one graph instead of two, i.e. to assume that $G=H$.

\begin{definition} Internal solutions for a graph $G$ are vectors $(X,Y, Z, T)$
in $\sigma(G, G)$.
\end{definition}

\subsection{$\Lambda(G,G)$}

The orthogonality assumption, equation (\ref{equ2}), in the case
of $G=H$ can be written more efficiently. Indeed, it is easy to
check that (\ref{equ2}) is equivalent to
\begin{equation}
\langle X, g(i)\times g(j) \rangle=(Y(j)-Z(i))g_{ij},\ for\ every
\ i\neq j, \label{equ3}
\end{equation}
\begin{equation}
 \langle X, g(i)\times g(i) \rangle=T(i),\ for\ every\ i.
 \label{equ4}
\end{equation}

\vspace{4mm}

\begin{lem}\label{ai}{ Assume that the graph $G$ is connected. Then, for every
$(X, Y, Z, T)$ $\in \Lambda (G, G)$, the function $Y+Z$ on the
vertices is constant, i.e., $Y+Z=aI$ for some number $a$.

 }\end{lem}

\begin{proof}{ Assume that $i, j$ are two adjacent vertices
in $G$. The condition (\ref{equ3}) implies that $$\langle X,
g(i)\times g(j) \rangle=(Y(j)-Z(i))g_{ij},$$ $$\langle X,
g(i)\times g(j) \rangle=(Y(i)-Z(j))g_{ji}.$$ Therefore
$Y(i)+Z(i)=Y(j)+Z(j)$, for any two adjacent vertices.
Connectivity of $G$ implies the desired conclusion.

}
\end{proof}

From now on, we assume that $G$ is a connected graph. Lemma
\ref{ai} gives us a partitioning of the set $\Lambda(G,G)$, as
follows:
$$\Lambda_a(G, G)=\{(X, Y, Z,
T)\in \Lambda(G, G): Y+Z=aI\}.$$ Notice that, for any $a\in
\mathbf{F}_q$, $$(0, aI, aI, 0)\in \Lambda_{2a}(G, G),$$ and
$$\Lambda_{2a} (G, G)=\Lambda_0(G, G)+(0, aI, aI, 0).$$ Therefore,
since $q$ is an odd number, all $\Lambda_a(G, G)$'s are just
shifts of $\Lambda_0 (G, G)$.

\vspace{4mm}

\begin{definition} Consider a connected graph, $G$. We say that $ij$ is an edge of $G$
if $g_{ij}\neq 0$.  For an even cycle $C$ in $G$ consisting of
(ordered) vertices $i_1, i_2, \cdots i_{2l}$, let
$$v(C)=\sum^{2l}_{k=1} (-1)^{k}g_{i_k i_{k+1}}^{-1} g(i_k)\times
g(i_{k+1}).$$ We define $\nu (G)$, called the {\it bineighborhood
space} of $G$, to be the subspace generated by the vectors
$g(i)\times g(j)$ for $i,j$ satisfying $g_{ij}=0$, as well as by
$v(C)$'s for even cycles $C$, i.e.,
$$\nu (G) = Span \{ \{v(C):\ C\ even\ cycle\}\cup\{g(i) \times g(j):\ g_{ij}=0\} \}.$$
\end{definition}

\vspace{4mm}

\begin{thm}\label{pen}{ For $X\in \mathbf{F}^{n}_q$, there exist
$Y, Z$ and $T$ such that $(X, Y, Z, T)\in \Lambda_0(G, G)$ if and
only if $X$ is orthogonal to $\nu (G).$ Moreover, if $G$ has an
odd cycle, for every $X\in \nu(G)^{\perp}$, there exists a unique
$(X, Y, Z, T)$ in $\Lambda_0(G,G)$.

 }\end{thm}

\begin{proof}{ Fix a vector $X$, and assume that there exists some $(X, Y, Z, T)\in \Lambda_0(G,
G).$ For any two vertices $i, j$ with $g_{ij}=0$,  $\langle X,
g(i)\times g(j) \rangle=(Y(i)+Y(j))g_{ij}=0$. Moreover, if $i_1,
i_2,\cdots, i_{2l}$ is a cycle then we have $$\langle X,
(-1)^kg^{-1}_{i_k i_{k+1}}g(i_k)\times g(i_{k+1})
\rangle=(-1)^k(Y(i_k)+Y(i_{k+1})).$$ By summing up all of these
equalities for $k=1, 2, \cdots, 2l$ we get that $X$ is orthogonal
to $\nu (C)$, for every even cycle $C$, and hence, to the whole
space $\nu (G)$.

For the other direction, suppose that $X\in \nu (G)^{\perp}$. For
any vertex $i$, set $T(i)=\langle X, g(i)\times g(i) \rangle$, and
$Z=-Y$. In the case that $G$ has an odd cycle, $Y$ would be
uniquely determined by (\ref{equ3}) on the vertices of that odd
cycle. Since, $G$ is connected, then one can determine the
function $Y$ on the rest of the vertices, and since $X\in \nu
(G)^{\perp}$ there is no ambiguity in the definition of $Y$.

In the case that $G$ contains no odd cycles, we can fix $Y(i)$
for some arbitrarily chosen vertex $i$, and then determine the
other components in terms of $Y(i)$. Once again, since $X\in \nu
(G)^{\perp}$ and there is no odd cycle, there is no ambiguity in
its definition.

}
\end{proof}

\subsection{Vectors in $\Lambda(G, G)$ have constant determinant}

Even though, the determinant is a quadratic function and not a
linear one, but in this setting, the set $\Lambda (G,G)$
satisfies some property that helps us to study this set more
deeply.

The problem of verifying locally equivalent graphs on one or two
vertices is a trivial problem, and hence, from now on we assume
that the number of vertices of $G$ is more than $2$.

\begin{thm}\label{det}{ For every $\phi \in \Lambda(G, G)$,
the determinant of $\phi$ is constant.

 }\end{thm}

\begin{proof}{ First, notice that one may restrict oneself to the case $\phi \in \Lambda_0 (G,G)$,
since there exists a vector $(X, Y, -Y, T)\in \Lambda_0 (G,G)$
and $a\in \mathbf{F}_q$ such that $\phi= (X, Y, -Y, T)+ a(0, I, I,
0)$, and $det\, \phi= det (X, Y, -Y, T)+a^2I.$ Hence, suppose that
$\phi =(X, Y, -Y, T)\in \Lambda_0(G, G).$

If $G$ contains at most two vertices, the proof is clear. So,
assume that  $n\geq 3$, and $i_1, i_2$ are two adjacent vertices
in $G$. Showing that $(det\, \phi)_{i_1}=(det\, \phi)_{i_2}$
gives us the desired result. By lemma \ref{tfb}, local
complementing operations, $\ast$ and $\circ$, do not change the
determinant, therefore we can assume that $i_1$ and $i_2$ have a
common neighbor $j_0$ by applying one $\ast$ operator if needed.
One has
$$g_{rs}(Y(r)+Y(s))=\langle X, g(r)\times g(s) \rangle,$$ for each
pair of unequal $r,s \in \{j_0, i_1, i_2\}$. Consequently,

$$Y(i_1)=\frac{1}{2}\Big[g_{i_1 i_2}^{-1}\langle X, g(i_1)\times
g(i_2)\rangle+g_{i_1 j_0}^{-1}\langle X, g(i_1)\times
g(j_0)\rangle -g_{i_2 j_0}^{-1}\langle X, g(i_2)\times
g(j_0)\rangle\Big],$$
and

$$Y(i_2)=\frac{1}{2}\Big[g_{i_1 i_2}^{-1}\langle X, g(i_1)\times
g(i_2)\rangle+g_{i_2 j_0}^{-1}\langle X, g(i_2)\times
g(j_0)\rangle -g_{i_1 j_0}^{-1}\langle X, g(i_1)\times
g(j_0)\rangle\Big].$$
On the other hand, $T(r)=\langle X,
g(r)\times g(r)\rangle$, for any vertex $r$. Hence, in order to
prove $(det\, \phi)_{i_1}=(det\,\phi)_{i_2}$, we should show that
$$-\frac{1}{4}\Big[g_{i_1 i_2}^{-1}\langle X, g(i_1)\times
g(i_2)\rangle+g_{i_1 j_0}^{-1}\langle X, g(i_1)\times
g(j_0)\rangle$$ $$-g_{i_2 j_0}^{-1}\langle X, g(i_2)\times
g(j_0)\rangle\Big]^2-X(i_1)\langle X, g(i_1)\times g(i_1)\rangle$$
$$= -\frac{1}{4}\Big[g_{i_1 i_2}^{-1}\langle X, g(i_1)\times
g(i_2)\rangle+g_{i_2 j_0}^{-1}\langle X, g(i_2)\times
g(j_0)\rangle$$ $$-g_{i_1 j_0}^{-1}\langle X, g(i_1)\times
g(j_0)\rangle\Big]^2-X(i_2)\langle X, g(i_2)\times
g(i_2)\rangle,$$
or equivalently
$$g_{i_1 i_2}^{-1}\langle X, g(i_1)\times g(
i_2)\rangle\Big[g_{i_2  j_0}^{-1}\langle X, g(i_2)\times
g(j_0)\rangle-g_{i_1 j_0}^{-1}\langle X, g(i_1)\times
g(j_0)\rangle\Big]$$
$$= X(i_1)\langle X, g(i_1)\times
g(i_1)\rangle-X(i_2)\langle X, g(i_2)\times g(i_2)\rangle.$$

Let $$C_j=g_{i_2 j}^{-1}\langle X, g(i_2)\times g(j)\rangle-
g_{i_1 j}^{-1}\langle X, g(i_1)\times g(j)\rangle,$$ for any $j$
adjacent  to both $i_1$ and $i_2$. Since, $X$ is orthogonal to the
cycle $j_0, i_1, j, i_2$, for any $j$ adjacent to $i_1, i_2$, we
have $C_{j_0}=C_j$. On the other hand, if either $g_{i_1 j}$ or
$g_{i_2 j}$ is zero, then
$$g_{i_1 j}  g_{i_2 j}X(j)C_{j_0}=0,$$
and
$$X(j)(g_{i_1 j}\langle X, g(i_2)\times g(j)
\rangle-g_{i_2 j}\langle X, g(i_1)\times g(j)\rangle)=0,$$
because, for instance if $g_{i_1 j}=0$, then $X$ is orthogonal to
$g(i_1)\times g(j).$ Therefore, we have

\begin{eqnarray*}& & \langle X, g(i_1)\times g(i_2)\rangle\Big[g_{i_2
j_0}^{-1}\langle X, g(i_2)\times g(j_0)\rangle- g_{i_1
j_0}^{-1}\langle X, g(i_1)\times g(j_0)\rangle\Big]\\
= & &\Big(\sum_{j\neq i_1,i_2} g_{i_1 j} g_{i_2
j}X(j)\Big)\Big[g_{i_2 j_0}^{-1}\langle X, g(i_2)\times
g(j_0)\rangle - g_{i_1 j_0}^{-1}\langle X, g(i_1)\times
g(j_0)\rangle\Big]\\
= & &\sum_{j\neq i_1,i_2} g_{i_1 j} g_{i_2 j}X(j)\Big[g_{i_2
j_0}^{-1}\langle X, g(i_2)\times g(j_0)\rangle - g_{i_1
j_0}^{-1}\langle X, g(i_1)\times g(j_0)\rangle\Big]\\
= & &\sum_{j\neq i_1,i_2} g_{i_1 j} g_{i_2 j}X(j)\Big[g_{i_2
j}^{-1}\langle X, g(i_2)\times g(j)\rangle - g_{i_1
j}^{-1}\langle X, g(i_1)\times g(j)\rangle\Big]\\
= & &\sum_{j\neq i_1,i_2} X(j) \Big[g_{i_1 j}\langle X,
g(i_2)\times g(j) \rangle-g_{i_2 j}\langle X, g(i_1)\times g(j)
\rangle\Big]\\
= & &\sum_{j\neq i_1,i_2} \sum^{n}_{k=1} g_{i_1
j}g_{i_2k}g_{kj}X(j)X(k)-g_{i_2 j} g_{i_1k}g_{kj}X(j)X(k)\\
= & & 0+\sum_{j\neq i_1,i_2} \sum_{k=i_1,i_2} g_{i_1 j} g_{i_2 k}
g_{k j} X(j)X(k)-g_{i_2 j} g_{i_1 k} g_{k j} X(j)X(k)\\
= & & \sum^{n}_{j=1} g_{i_1 i_2}g_{i_1 j}^2
X(i_1)X(j)-\sum^{n}_{j=1} g_{i_1 i_2} g_{i_2 j}^2 X(i_2)X(j)\\
= & & g_{i_1 i_2}\Big[X(i_1)\langle X, g(i_1)\times g(i_1)
\rangle-X(i_2)\langle X, g(i_2)\times g(i_2) \rangle\Big],
\end{eqnarray*}

which completes the proof.

}\end{proof}

\section{Linearity of the kernel of {\it det} function }

Using theorem \ref{det}, we may give another partition of the set
$\Lambda (G,G)$ as follows, $$\Lambda^\alpha (G, G)=\{\phi \in
\Lambda(G, G) : det\, \phi =\alpha I \},$$ and combining with the
previous partition, we set $$\Lambda ^\alpha_a(G,
G)=\Lambda_a(G,G)\cap \Lambda^\alpha (G, G).$$

Notice that $\Lambda_0$ is a linear subspace since it is the
kernel of a linear map. But generally the determinant function is
not a linear function when $q$ is not a power of 2. Here, due to
the nature and strength of theorem \ref{det}, we will show that
despite of not being a linear function, the kernel of the
determinant exhibit some linear properties. More precisely, we
show that $\Lambda^0_0(G,G)$ is a linear subspace if $dim\,
\Lambda (G, G)\geq 5$.

\subsection{Some useful lemmas}

For any $$\phi=(X, Y, -Y, T),\ \phi'=(X',Y', -Y', T')\in
\Lambda_0(G, G),$$ define
$$\Psi(\phi, \phi')=2Y\times Y'+X\times T'+ X'\times T.$$
Notice that, $\Psi (\phi, \phi')$ is constant, because $det$ is a
constant function and $$\Psi (\phi, \phi')=det\, \phi+det\,
\phi'-det\, (\phi+\phi').$$ Also, to prove the linearity of
$\Lambda^0_0(G,G)$, one can sufficiently show that $\Psi (\phi,
\phi')=0$ for any $\phi, \phi' \in \Lambda^0_0(G,G)$. The
following series of lemmas provide us the necessary tools.

\begin{lem}\label{si1}{ Suppose that $\phi_i=(X_i, Y_i, -Y_i, T_i)\in
\Lambda_0(G,G)$ for $i=1,2$, and moreover, $\phi_1\in
\Lambda^0_0(G,G)$. Also, suppose that $\Psi (\phi_1,\phi_2)=aI$
for some $0\neq a\in \mathbf{F}_q$. Then on every vertex, either
$X_1$ or $X_2$ is non-zero.

}\end{lem}
\begin{proof}{ Assume that $X_1(i)=0,$ for some vertex $i\in
\{1,2,\cdots,n\}$. Since $det\, \phi_1=0$, one has
$-Y_1(i)^2-X_1(i)T_1(i)=0$ and therefore, $Y_1(i)=0$. Hence, the
$i$-th component of $\Psi (\phi_1, \phi_2)$ is $X_2(i)T_1(i)=a
\neq 0$ and thus $X_2(i)\neq 0.$

}
\end{proof}

\begin{lem}\label{si2}{ Let $\phi_i=(X_i, Y_i, -Y_i, T_i)\in \Lambda^0_0(G,G)$ for $i=1,2$,
and $\psi=(U,V, -V,W)$ $\in \Lambda^0_0(G,G)$. Moreover, assume
that $\Psi (\phi_1,\phi_2)=aI$ for some $0\neq a\in
\mathbf{F}_q$, and $\Psi(\phi_1, \psi)=0$. Then $supp(U)\subseteq
supp(X_1)$, in the sense that if $U$ is non-zero on some vertex,
then so is $X_1$.

 }\end{lem}

\begin{proof}{ For any $r\in \mathbf{F}_q$, we have $\Psi (\phi_1, r\psi+\phi_2)=aI\neq
0$. Therefore, by lemma \ref{si1}, $supp(X_1)\cup
supp(rU+X_2)=\{1,2, \cdots, n\}.$ Now suppose that $X_1(i)=0$ for
some $i\in \{1,2,\cdots, n\}$, and $U(i)\neq 0$. There exists some
$r_0\in \mathbf{F}_q$ such that $r_0U(i)+X_2(i)=0.$ Thus $i\notin
supp(X_1)\cup supp(r_0U+X_2)$, which contradicts the earlier
statement.

}
\end{proof}

The third lemma is the following:

\begin{lem}\label{si3}{ Suppose that $\phi_i=(X_i, Y_i, -Y_i, T_i)\in
\Lambda^0_0(G,G)$ for $i=1,2$, such that $supp(X_1)$ and
$supp(X_2)$ are minimal subsets of $\{1,2, \cdots ,n\}$ with $\Psi
(\phi_1, \phi_2)\neq 0$. If $\psi=(U,V, -V, W)\in
\Lambda^0_0(G,G)$ and $\Psi (\phi_1, \psi)=0$, then either
$\psi=0$ or $U=-a^{-1}c X_1$, where $\Psi(\phi_1, \phi_2)=aI$ and
$\Psi(\phi_2, \psi) =cI.$

 }\end{lem}

\begin{proof}{ First, assume that $c=0$. In this case we have $\Psi(r\psi+\phi_1,
\phi_2)=aI$, for any $r\in \mathbf{F}_q$. If $U(i)\neq 0$ for some
$i\in \{1,2, \cdots n\}$, then there exists $r_0\in \mathbf{F}_q$
such that $r_0U(i)+X_1(i)=0$. Thus, using lemma \ref{si2},
$supp(U)\subseteq supp(X_1)$ and hence $supp(r_0U+X_1)$ is a
proper subset of $supp(X_1)$. Moreover, $\Psi(r_0\psi+\phi_1,
\phi_2)=aI$ and $det(r_0\psi+\phi_1)=r^2_0 det\ \psi+det\
\phi_1-r_0\Psi(\psi, \phi_1)=0$. Then $r_0\psi+\phi_1\in
\Lambda^0_0(G,G)$, which contradicts the minimality of $\phi_1,
\phi_2$. Therefore $U=0$.

Now, assume that $c\neq 0$. Once again, by lemma \ref{si2},
$supp(U)\subseteq supp(X_1)$ and $supp(X_1)\subseteq supp(U)$.
Suppose that $U(i)\neq 0$ for some $i\in \{1, \cdots, n\}$. There
exists some $r_0\in \mathbf{F}_q$ such that $r_0U(i)+X_1(i)=0$,
and also, $\Psi (r_0\psi+\phi_1, \phi_2)=(r_0c+a)I$. By the
minimality of $\phi_1$ and $\phi_2$, one concludes that
$r_0c+a=0$. Therefore, $-ac^{-1}U(i)+X_1(i)=0$ for any $i$
satisfying $U(i)\neq 0$. Therefore $U=-a^{-1}c X_1$.

}
\end{proof}

Finally, the last lemma is a fact in number theory.

\begin{lem}\label{33}{ Given a $3 \times 3$ matix $A$
in $\mathbf{F}_q$, there exists a non-zero vector ${\bf{x}} \in
\mathbf{F}_q ^3$ so that ${\bf{x}}^T \cdot A \cdot {\bf{x}} =0$.}
\end{lem}

\begin{proof}{There are many ways to prove this lemma, and one
straight forward computational way is as follows. Let us rewrite
the matrix equation as a degree two numeric equation,
$$ax^2+by^2+cz^2+2dxy+2exz+2fyz=0.$$
If $abc=0$ then there exists a trivial solution to the equation.
For instance, $(1,0,0)$ when $a=0$. Thus, we assume that $abc\neq
0$.

Solving the latter equation in terms of $z$ using {\it the square
root of delta} formula, we obtain that the problem is equivalent
to this one: {\it does delta have a square root?} In other words,
it is equivalent to finding a non-trivial solution to the
following equation,
$$\alpha x^2 + \beta y^2+2\gamma xy =t^2,$$
where $\alpha=e^2-ac, \beta=f^2-bc$ and $\gamma =ef-cd$. Once
again, if $\alpha=0$ then $(x,y,t)=(1,0,0)$ is a solution, and if
not, by solving it in terms of x, we get the next equation,

$$\theta y^2=s^2-\alpha t^2,$$ where $\theta=\gamma^2-\alpha\beta$.
We set $y=1$. For different values of $s,t \in \mathbf{F}_q$, each
of the functions $s^2-\theta$ and $\alpha t^2$ ranges over
$(q+1)/2$ different elements. Therefore, there is at least one
$(s,t)$, such that $s^2-\theta=\alpha t^2$, and we are done.

}\end{proof}

\subsection{$\Lambda^0_0(G,G)$ is linear}

Now, we have all the necessary tools in hand, to provide a proof
for the linearity of $\Lambda^0_0(G,G)$.

\begin{thm}\label{si}{ Suppose that $dim\, \Lambda_0 (G,G)\geq 5$. Then $\Psi\equiv 0$
on $\Lambda^0_0(G,G),$ or equivalently, $\Lambda^0_0(G,G)$ is a
linear subspace.
 }\end{thm}

\begin{proof}{ First of all, we can assume that $G$ has an odd cycle.
Because we already know that by local complementation, the linear
properties of $\Lambda(G,G)$, the determinant and so the function
$\Psi$, do not change. Under this assumption, as we observed in
theorem \ref{pen}, if $\phi_i=(X_i,Y_i, -Y_i, T_i)\in
\Lambda_0(G,G)$, $i=1,2$, and $X_1=X_2$, then $\phi_1=\phi_2$.

Suppose that, $\Psi$ is not zero, and let $\phi_i=(X_i, Y_i, -Y_i,
T_i)\in \Lambda^0_0(G,G)$, $i=1,2$, such that $X_1, X_2$ are
minimal elements of $\{1,2,\cdots, n\}$ (in the sense of lemma
\ref{si3}) satisfying $\Psi (\phi_1, \phi_2)=aI$, where $0\neq
a\in \mathbf{F}_q$. Since, $dim\, \Lambda_0(G,G)\geq 5$, there
exist elements $\psi_j=(U_j, V_j, -V_j, W_j)\in \Lambda_0 (G, G)$,
$j=1,2,3$, independent of $\phi_1$ and $\phi_2$. Set
$\Psi(\phi_1, \psi_j)=b_j$ and $\Psi(\phi_2, \psi_j)=c_j$ for
$j=1,2,3$, and also define
$\omega_j=a\psi_j-c_j\phi_1-b_j\phi_2$, for $j=1,2,3.$  One can
easily verify that $\Psi (\phi_i, \omega_j)=0$, for every $i=1,2$
and $j=1,2,3$. Using lemma \ref{33}, we can find a non-trivial
solution of $det\, (r_1\omega_1+r_2\omega_2+r_3\omega_3)=0$,
where $r_1,r_2,r_3\in \mathbf{F}_q$. Thus,
$r_1\omega_1+r_2\omega_2+r_3\omega_3\in \Lambda^0_0(G, G)$ and
$\Psi(\phi_i, r_1\omega_1+r_2\omega_2+r_3\omega_3)=0,$ $i=1,2$.
Therefore, by lemma \ref{si3} the first coordinate of
$r_1\omega_1+r_2\omega_2+r_3\omega_3$ is zero and also by theorem
\ref{pen}, we conclude that
$r_1\omega_1+r_2\omega_2+r_3\omega_3=0$, which is a
contradiction. Hence, $\Psi (\phi_1, \phi_2)=0$ for every
$\phi_1,\phi_2 \in \Lambda^0_0(G,G),$ and $\Lambda^0_0(G,G)$ is a
linear subspace.

}\end{proof}

This theorem says that $\Lambda^0_0(G,G)$ is a linear subspace.
But, we can say much more about that. Having constant determinant
as well as its linearity, makes it a significantly helpful to
study $\sigma (G,G)$ whose description is our main goal in this
section. In fact, the following lemmas tell us that this linear
space, $\Lambda_0 ^0(G,G)$, is really a {\it large} subspace in
the whole space $\Lambda (G,G)$.

\begin{lem}{ The co-dimension of $\Lambda_0 ^0(G,G) $ in $\Lambda_0(G,G)$ is at most two, provided
that  $dim\, \Lambda_0 (G,G) \geq 5$ .}
\end{lem}

\begin{proof}{ Consider three independent vectors $ \phi_1,\phi_2,\phi_3 \in
 \Lambda_0(G,G)$. For numbers $c_1,c_2$ and $c_3$,
$$ det\, (c_1 \phi_1+c_2 \phi_2+c_3 \phi_3) = c_1^2 det\,(\phi_1 )+c_2^2 det\,(\phi_2 )+
c_3^2 det\,(\phi_3 )$$
$$+c_1 c_2 \Psi(\phi_1 ,\phi_2 )+c_1 c_3 \Psi(\phi_1 ,\phi_3 )+
c_2 c_3 \Psi(\phi_2 ,\phi_3 ).$$

By lemma \ref{33}, there exists $(c_1,c_2,c_3) \neq 0$ such that
$det\, (c_1 \phi_1+c_2 \phi_2+c_3 \phi_3) = 0$, which means that
$c_1 \phi_1+c_2 \phi_2+c_3 \phi_3 \in \Lambda^0_0(G,G)$. Thus, the
co-dimension of $\Lambda_0 ^0(G,G)$ inside $\Lambda_0(G,G)$ is at
most two.

}\end{proof}

Since $q$ is an odd number, the translation of $\Lambda_0(G,G)$
by vectors $(0,aI,aI,0)$, for different values $a \in
\mathbf{F}_q$, gives the whole space $\Lambda (G,G)$. Therefore,
co-dimension of $\Lambda_0(G,G)$ in $\Lambda(G,G)$ is one. Also,
$\Lambda^0_0(G, G)+(0,I,I,0)=\Lambda^1_2(G,G)$. On the other
hand, by the definition of $\sigma$, $\sigma(G,G)\supseteq
 \Lambda^1_2(G,G).$ Therefore, we conclude the following corollary:

\begin{cor}\label{cor1} {There exists an affine linear subspace inside $\sigma (G,G)$, whose co-dimension
in the whole space $\Lambda (G,G)$ is at most $3$, provided that
$\Lambda (G,G)$ has dimension not less than $5$. Putting these
together, in general the co-dimension of this affine subspace is
at most $5$.}
\end{cor}

Having the mentioned property in hand, together with the $\beta$
function introduced earlier, we can now give the desired
description of $\sigma(G,H)$ for equivalent graphs $G$ and $H$,
which creates the foundations of our algorithm determining
whether two graphs are equivalent or not.

\begin{thm}\label{cor2}{ If the connected graphs $G$ and $H$, defined on the same vertex sets, are locally
equivalent, then the co-dimension of some affine linear subset of
$\sigma (G,H)$ inside $\Lambda (G,H)$ is at most $5$. }
\end{thm}

\section{The algorithm}

We now have all the tools to describe in details and provide the
proof for an efficient algorithm to determine whether two graphs
are equivalent or not. The algorithm is the following.

Suppose that $G$ and $H$ are two connected graphs (notice that by
local complementation a connected graph remains connected), with
neighborhood functions $g$ and $h$. Consider the linear system of
equations:

\begin{equation}
\langle X, g(i)\times h(j)\rangle-\langle Y, g(i)\times
e_j\rangle+\langle Z, e_i\times h(j)\rangle-\langle T, e_i\times
e_j\rangle=0, \label{equ5}
\end{equation}
for any two vertices $i, j$, and for $X,Y,Z$ and $T$ in
$\mathbf{F}_q ^n$, together with the equation
\begin{equation}
Y\times Z-X\times T = I. \label{equ6}
\end{equation}
Assume that $\mathcal{B}$ is an arbitrary basis for
$\Lambda(G,H)$, the set of solutions of the linear equation
(\ref{equ5}), which can be computed efficiently. According to the
corollary \ref{cor1} and theorem \ref{cor2} in the previous
section, if there exist solutions for (\ref{equ5}) and
(\ref{equ6}), then there exists an affine subset, denoted by
$\Gamma$, in $\Lambda(G,H)$ with $codim \leq 5$, whose elements
all satisfy both (\ref{equ5}) and (\ref{equ6}). The following
lemma takes the advantage of this property.

\begin{lem}\label{final} { For any basis $\mathcal{B}$ of a linear space $\Lambda$, and every affine
subspace $\Gamma$ of $\Lambda$ of $codim \leq 5$, there exists a
vector $u \in \Gamma$, which is a linear combination of at most
five elements of $\mathcal{B}$.}
\end{lem}

\begin{proof}{ Consider the set $\Gamma' = \{ u - v : u,v \in \Gamma\}$ which is a
subspace of $\Lambda$, and the {\it canonical projection} $p:
\Lambda \rightarrow \Lambda / \Gamma'$. The set $p(\mathcal{B})$
generates $\Lambda / \Gamma'$, and therefore there exists a basis
 $\{p(b_1), \dots, p(b_k)\}$ for $\Lambda / \Gamma'$, where $b_1,\dots b_k\in
 \mathcal{B}$ and
 $k=dim\,(\Lambda) - dim\, (\Gamma') \leq 5.$ Since $\Gamma$ is affine, $\Gamma \in {\Lambda / \Gamma'}$
 and hence can be written as the linear combination of $p(b_i)$'s, which means that there exists
a vector $u \in \Gamma$ which is a linear combination of at most
five elements of $\mathcal{B}$.

}\end{proof}

By this lemma, we can now consider all of the linear combinations
of every $5$ elements of $\mathcal{B}$, and check whether or not
it satisfies the condition (\ref{equ6}). If at least one of them
satisfies (\ref{equ6}), then the answer is {\it positive}, and is
{\it negative} otherwise.

Notice that, solving this problem for {\it disconnected} graphs
is an immediate consequence of solving it for the connected
graphs, since the local operators preserve the connectivity.

The described algorithm is efficient. In fact, by using a
pivoting method, a basis $\mathcal{B}$ can be computed in $O(n^4)$
time, because there are $O(n^2)$ linear equations in (\ref{equ5}).
This number must be added to and hence will be dominated by the
time to check the equation (\ref{equ6}) for all of the linear
combination of five elements of this basis, which is $O(n^5)$,
(in the case that $dim \Lambda (G,H) \leq 5$, we check all of the
possibilities). Thus, the algorithm takes $O(n^5)$ time, and the
overall complexity is polynomial in $n$.

\vspace{4mm}

\noindent{\bf Acknowledgement.} Authors are greatly thankful to
Prof. Peter W. Shor, for all his gracious support and helpful
advice. They are also thankful to Prof. Isaac Chuang for
introducing this problem, and for all useful comments he kindly
gave them.

\end{document}